\documentstyle[sprocl]{article}

\input{psfig}
\def\Journal#1#2#3#4{{#1} {\bf #2}, #3 (#4)}

\def\NPB{{\em Nucl. Phys.} B}
\def\PLB{{\em Phys. Lett.}  B}

\def\PRD{{\em Phys. Rev.} D}

\def\ARNPS{\em Ann. Rev. Nucl. Part. Sci.}
\def\PU{\em Phys. Usp.}
\def\JL{\em JETP Lett.}
\def\simlt{\stackrel{<}{{}_\sim}}
\def\simgt{\stackrel{>}{{}_\sim}}

\def\be{\begin{equation}}
\def\ee{\end{equation}}
\def\bea{\begin{eqnarray}}
\def\eea{\end{eqnarray}}

\begin{document}
\begin{titlepage}
\title{
PERTURBATIVE ANALYSIS OF THE MSSM ELECTROWEAK PHASE TRANSITION}
\vspace{1cm}
\author{ J.R. ESPINOSA \\
Department of Physics and Astronomy, University of Pennsylvania\\
Philadelphia, PA 19104-6396, USA}

\maketitle
\vspace{1.cm}
\def\baselinestretch{1.15}
\begin{abstract}
Light stops can give a strong first order electroweak phase transition
in the Minimal Supersymmetric Standard Model (MSSM) satisfying one of the
necessary conditions for electroweak baryogenesis. I discuss the region
of parameter space where this occurs using a perturbative analysis of the
Higgs effective potential. Two-loop QCD corrections associated
with stop loops are crucial to open this baryogenesis window,
which corresponds to $2.25\simlt \tan\beta\simlt 3.6$, one stop not much
heavier than the top quark, $m_A>120\ GeV$ and a light Higgs boson with 
$m_h< 85\ GeV$. This region will be explored by LEP~II very soon.
\end{abstract}
\vspace{4cm}
\leftline{June 1997}

\thispagestyle{empty}

\vskip-15.cm
\rightline{{ UPR--0758-T}}
\rightline{{ IEM--FT--158/97}}
\rightline{{ hep-ph/9706389}}

\vskip3in

\end{titlepage}

\title{
PERTURBATIVE ANALYSIS OF THE MSSM ELECTROWEAK PHASE TRANSITION}

\author{ J.R. ESPINOSA }

\address{
Department of Physics and Astronomy\\
University of Pennsylvania\\
Philadelphia, PA 19104-6396, USA}

\maketitle\abstracts{
Light stops can give a strong first order electroweak phase transition 
in the Minimal Supersymmetric Standard Model (MSSM) satisfying one of the
necessary conditions for electroweak baryogenesis. I discuss the region
of parameter space where this occurs using a perturbative analysis of the
Higgs effective potential. Two-loop QCD corrections associated
with stop loops are crucial to open this baryogenesis window,
which corresponds to $2.25\simlt \tan\beta\simlt 3.6$, one stop not much 
heavier than the top quark, $m_A>120\ GeV$ and a light Higgs boson with 
$m_h< 85\ GeV$. This region will be explored by LEP~II very soon.}
  

The study of the properties of the electroweak phase transition (and in particular
of its strength) in the MSSM is motivated by the possibility of electroweak
baryogenesis~\cite{review} (which cannot take place in the minimal Standard
Model). A necessary
condition for a successful baryogenesis at the electroweak phase transition is
that this transition is strong enough to suppress sphaleron reactions in the
broken phase, preventing the erasure of the created baryon asymmetry.
In this talk, I address the question of whether this constraining condition can be
met in some region of the MSSM parameter space (improving thus over the situation
in the SM). The emphasis here will be on the qualitative aspects rather than in
the quantitative details, which can be found in refs.~\cite{dce}.

To a great extent, the prospects for a strong transition depend on the details of
the $T=0$ Higgs potential. The MSSM Higgs sector contains two Higgs doublets
$H_{1,2}$ of opposite hypercharge. Electroweak breaking is described by a Higgs
potential which is a function of two fields $\varphi_{1,2}$ ($\sim Re H_{1,2}^0$).
In the broken phase, the Higgs spectrum consists of two neutral scalars $h^0, H^0$
($m_h<m_H$), one pseudoscalar $A^0$ and a charged Higgs pair $H^\pm$ and, at
tree-level, the properties of the Higgs sector are determined by
$\tan\beta=\langle H^0_2\rangle/\langle H^0_1\rangle=v_2/v_1$ and $m_A$ (with
$v^2=v_1^2+v_2^2= (246\ GeV)^2$ fixed by the gauge boson masses).

In the $(\varphi_1,\varphi_2)$ plane, the minimum of the $T=0$ potential lies
along
the direction determined by $\tan\beta$. When $m_A\gg m_Z$, the mass eigenstate
$h^0$ (which always has a mass controlled by the electroweak scale) is aligned
with that breaking direction while the orthogonal $H^0$ direction
has mass $m_H\simeq m_A$. In that case, the low-energy potential reduces to a
one-dimensional SM-like potential $V(h^0)$. The same is true for the
temperature dependent potential as long as $T\ll m_H$. For $m_A\sim m_Z$, $h^0$
forms some angle with the breaking direction and the full two Higgs potential
should be considered.
This picture is not affected qualitatively by radiative corrections which,
however, change sizably
the quantitative properties of the Higgs sector. The main effect is that the mass
along the breaking direction is
\begin{equation}
m_\beta^2=m_Z^2\cos^22\beta
+\frac{3}{2\pi^2}\frac{m_t^4}{v^2}\log\frac{m_{\tilde{t}_1}m_{\tilde{t}_2}}{m_t^2},
\end{equation}
where the second term comes from top-stop (with masses $m_t=h_t\varphi_2$ and
$m_{\tilde{t}_{1,2}}$ respectively) loops. The mass of $h^0$ satisfies $m_h\leq
m_\beta$.

The temperature dependent effective potential can be computed up to two-loops
in perturbation theory (with resummation of important thermal effects)
using standard techniques~\cite{effpot}, and the electroweak transition at $T_c$
studied. To a good
approximation, this transition proceeds along a fixed direction
$\varphi_2/\varphi_1\sim \tan\beta_{T_c}=v_2(T_c)/v_1(T_c)$. The numerical
analysis shows that $\tan\beta_{T_c}\simgt \tan\beta$ ($\gg$ for small $m_A$ and
$\simeq$ for large $m_A$). The transition strength, as measured by $v(T_c)/T_c$, 
is controlled by the quartic Higgs coupling along the direction of breaking
\begin{equation}
\frac{v(T_c)}{T_c}\sim\frac{1}{\lambda_{\beta_{T_c}}},
\end{equation}
and $\lambda_{\beta_{T_c}}$ is directly proportional to $m^2_{\beta_{T_c}}$ 
(given by Eq.~(1) with $\beta\rightarrow \beta_{T_c}$) which, in general, is not
the mass of any physical Higgs boson. The condition to avoid the erasure of
baryons by sphalerons right after the transition is~\cite{shap} $v(T_c)/T_c>1$.

We can now describe how $v/T_c$ depends on the $(m_A,\tan\beta)$ parameters.
At large $m_A$ ($\gg m_Z,T_c$), $\beta_{T_c}\simeq \beta$ and
$m^2_{\beta_{T_c}}\simeq m_{h}^2$. From the $\tan\beta$ dependence of Eq.~(1) 
we conclude that $v/T_c$ is larger for $\tan\beta\simeq 1$ (which minimizes
$m_h$).
For small $m_A$ ($\sim m_Z,T_c$), $\beta_{T_c}\gg \beta$ and larger values of
$m^2_{\beta_{T_c}}$ are probed. In this case, having a light $h^0$ does not help
to
increase $v/T_c$
because $h^0$ lies along a direction which is not the breaking direction.
We conclude that, for fixed $\tan\beta$, $v/T_c$ is smaller for smaller $m_A$ and
the region where the transition is stronger is that of large
$m_A$ and small $\tan\beta$. In that region, $h^0$ (and its potential) is SM-like.
This means that, without extra SUSY contributions to the potential, the
SM result $v/T_c\ll 1$ for $m_h>70.7\ GeV$ (LEP II limit~\cite{ALEPH}) is
recovered.
Thus, the
two-doublet structure of the potential, being constrained by Supersymmetry, is of
no help to strengthen the transition.

The important SUSY effects that can change this situation are those of the stops,
which, being bosons, with a
large number of degrees of freedom and a large coupling ($\sim h_t$) to the Higgs
fields can affect $v/T_c$ sizably through contributions of infrared origin to the
potential. The field-dependent masses of the stops are of the form (neglecting
$\tilde{t}_L-\tilde{t}_R$ mixing which tends to weaken the transition and $D$
terms for simplicity)
\begin{equation}
m_{\tilde{t}}^2(\varphi,T)\sim \tilde{m}^2 + h_t^2\varphi_2^2 +
\Pi_{\tilde{t}}(T),
\end{equation}
where $\tilde{m}$ ($m_{Q,U}$ for $\tilde{t}_{L,R}$ respectively) are soft-susy
breaking masses and $\Pi_{\tilde{t}}\sim g_s^2T^2$ is a thermal contribution
to the effective stop masses from interactions with the surrounding plasma.
In the absence of $\tilde{m}^2$ and $\Pi_{\tilde{t}}$, static stop modes
contribute
to the 1-loop potential~\cite{one} a term $\sim -Th_t^3\varphi^3$ which enhances
$v/T_c$ significantly. At two-loops~\cite{dce} the same modes, in a 
$\tilde{t}-\tilde{t}-gluon$ setting-sun diagram, contribute $\sim -h_t^2g_s^2T^2
\varphi^2\log\varphi^2$ with a similar enhancement effect on $v/T_c$.
Non-zero $\tilde{m}^2$ and $\Pi_{\tilde{t}}$ screen this ideal behaviour reducing
the effect, so that the transition is stronger for smaller $m_Q^2$ and $m_U^2$. 
A lower limit on $m_Q^2$ can be obtained requiring that the contribution of the
$(\tilde{t},\tilde{b})_L$ doublet to the $\rho$ parameter stays within
experimental limits. There is no such constraint on $m_U^2$ so that we set it
to zero. Negative values of $m_U^2$ are associated with the existence of a 
dangerous color breaking minimum of the full scalar potential along the stop
direction. Modest negative values of $m_U^2$ are however still allowed
cosmologically and can further increase $v/T_c$.~\cite{cqw} Loops of stop
static modes have no infrared problems because $\tilde{m}^2$ and $\Pi_{\tilde{t}}$
cut-off infrared divergences. However, they can reapear if
$m_U^2+\Pi_{\tilde{t}}\sim 0$,
casting some doubt on the reliability of the perturbative analysis of the
$m_U^2<0$ region. To be on the safe side I restrict my discussion to the 
$m_U^2\simgt 0$ region. For analyses of the $m_U^2<0$ region, see
refs.~\cite{cqw}$^{\!,\,}$\cite{mu2lt0}.

\begin{figure}[hbt]
\psfig{figure=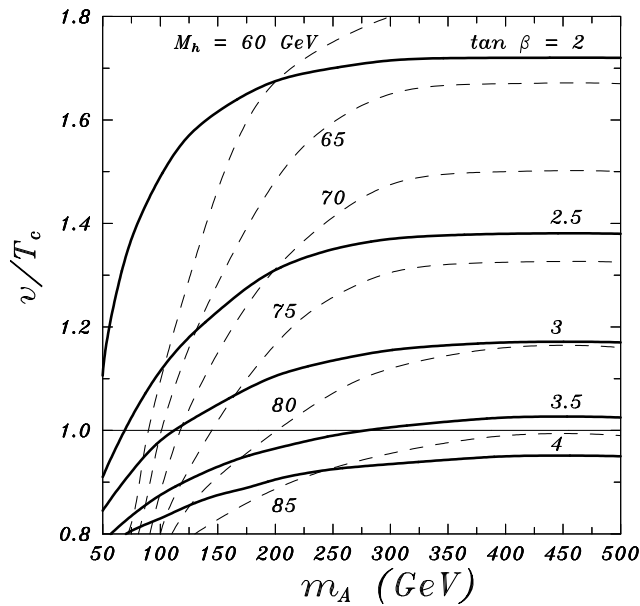,height=9cm,width=9cm,bbllx=0.cm,bblly=2.cm,bburx=13.cm,bbury=15cm}
\caption{$v/T_c$ (solid) for different values of $\tan\beta$ and $m_h$ (dashed) as a
function of the pseudoscalar mass $m_A$. $M_t=175\ GeV$, $m_Q=250\ GeV$ and
$m_U=0$.}
\end{figure}

Figure 1 contains the results for $v/T_c$ from an analysis~\cite{dce} of the
effective
potential up to two-loops in resummed perturbation theory with $M_t=175\ GeV$,
$m_Q=250\ GeV$, $m_U=0$ and zero stop mixing. It shows how $v/T_c$ can be larger
than 1 for
low values of $\tan\beta$ (solid lines) and large enough $m_A$. Very low
$\tan\beta$ corresponds to a light $h^0$ (lines of constant $m_h$ are dashed),
which is experimentally excluded.
Imposing the LEP limit ($m_h\simgt 70\ GeV$) we find a window for baryogenesis
which has
$2.25\simeq \tan\beta \simeq 3.6$, $m_A>120\ GeV$ and $m_h<85\ GeV$.
In addition, one of the stops should be not much heavier than $M_t$ (values
slightly bigger are still allowed). Comparison of this with 1-loop
results~\cite{one} shows the importance of the two-loop corrections to achieve
$v/T_c>1$. The large size of
two-loop contributions is not a symptom of the breaking of the
perturbative expansion because QCD corrections appear first at that order.
Similar 2-loop corrections were present in the SM for top loops,
but there no enhancement of $v/T_c$ was present due to the fermionic nature of the
top quark ($\delta V\sim h_t^2\varphi^2T^2$ in that case). 

In the SM, the resummed perturbative analysis of the electroweak phase transition
along the same lines is not reliable for values of the Higgs mass larger than 
$m_W$. One can estimate the loop expansion parameter (e.g. in
the vicinity of the broken minimum) as $\epsilon_{SM}\sim
g^2T/m_W(\varphi,T)\sim\lambda/g^2$ (in the SM, the transition is driven by
gauge boson loops. $\epsilon_{SM}$ is the cost of an extra $W$ loop). To have
$\epsilon_{SM}<1$ requires $m_h\simlt m_W$.
 In the MSSM, when stops drive the transition, the cost of an extra stop loop
is $h_t^2T/m_{\tilde{t}}(\varphi,T)\sim \lambda/h_t^2$ and we see that our
analysis is expected to be reliable for larger values of $m_h$. In particular, the
window found for baryogenesis falls within the region where our loop expansion is
under control.

Interesting alternative approaches~\cite{3d} to the study of the MSSM electroweak
phase transition have been followed using 3d reduced effective theories plus
lattice simulations (see talks by M. Laine and J. Cline) with qualitatively
similar results. However, these analyses are less reliable precisely in the
small $m_U^2$ region, which is the interesting one for baryogenesis. Clearly, more
work along these lines would be desirable.

To conclude, if stops are close to the Fermi scale (as originally motivated by
Supersymmetry) they drive the electroweak phase transition and can give $v/T_c>1$
as a necessary ingredient for electroweak baryogenesis. QCD radiative effects are
crucial to this point. Nevertheless, the available parameter space is very
constrained. In particular, the condition $m_h<85\ GeV$ will be challenged by
LEP~II very soon. Let us hope SUSY gets lucky once more.

\section*{Acknowledgments}
I thank B. de Carlos for an enjoyable collaboration on the topic presented. This
work was supported by the U.S. Department of Energy Grant No. DOE-EY-76-02-3071.

\end{document}